\documentclass[fleqn,twoside]{article}
\usepackage{gc}
\usepackage{amssymb,amsmath}

\heads{Vladimir V. Kassandrov}
      {Algebrodynamics over complex space and phase extension...}

\newcommand{\be}{\begin{equation}\label}
\newcommand{\ee}{\end{equation}}
\newcommand{\prt}{\partial}
\newcommand{\p}{\prime}

\def\B{{\mathbb B}}
\def\C{{\mathbb C}}
\def\Q{{\mathbb Q}}
\def\R{{\mathbb R}}

\begin{document}
\twocolumn[

\Title{ALGEBRODYNAMICS OVER COMPLEX SPACE \yy 
AND PHASE EXTENSION OF THE MINKOWSKI GEOMETRY}


   \Author{Vladimir V. Kassandrov\foom 1            
              }   
          {Institute of Gravitation and Cosmology, Peoples' Friendship University of Russia, 6 Mikluho-Maklay St., \\ Moscow 117198, Russia}              


\Abstract
    {First principles should predetermine physical geometry and  
dynamics both together. In the ``algebrodynamics'' they follow solely 
from the  properties of biquaternion algebra $\B$ and the analysis over $\B$. 
We briefly present the algebrodynamics over the Minkowski background based 
on a nonlinear generalization to $\B$ of the Cauchi-Riemann analyticity 
conditions. Further, we consider the effective real geometry uniquely 
resulting from the structure of multiplication in $\B$ and found it to be     
of the Minkowski type, with an additional phase invariant. Then we pass 
to study the primordial dynamics that takes place in the complex $\B$ space 
and brings into 
consideration a number of remarkable structures: an ensemble of 
identical correlated matter pre-elements (``duplicons''), caustic-like 
signals (interaction carriers), a concept of random complex time 
resulting in irreversibility of physical time 
at a macrolevel, etc. In partucular, the concept of ``dimerous electron'' 
naturally arises in the framework of complex algebrodynamics and, together 
with the above-mentioned phase invariant, allows for a novel approach to 
explanation of quantum interference phenomena alternative to the recently  
accepted paradigm of wave-particle dualism.}

]  
\email 1 {vkassan@sci.pfu.edu.ru}

\section{Status of Minkowski geometry and the algebrodynamical paradigm}

A whole century after German Minkowski introduced his famous conception of the 4D space-time continuum, 
we come to realize  the restricted nature of this conception and the necessity of 
its revision, supplement and derivation from some general and fundamental 
principle. 

Indeed, formalism of the 4D space-time geometry was indispensable to  
ultimately formulate the Special Theory of Relativity (STR), to ascertain  
basic symmetries of fundamental physical equations and related conservation 
laws. It was also the Minkowski geometry that served as a base for  
formulation of the concept of {\it curved\/} space-time in the   
framework of the Einstein's General Theory of Relativity (GTR).

Subsequently, Minkowski geometry and its pseudo-Riemannian analog have 
been generalized via introduction of {\it effective\/} geometries related to 
correspondent field dynamics (in the formalism of {\it fiber bundles\/}),  
or via exchange of Riemannnian manifold for spaces with torsion, non-metricity 
or additional ``hidden'' dimensions (in the {\it Kaluza-Klein 
formalism}). There have been considered also  the models of discrete 
space-time, and a challenging scheme of the {\it causal sets}~\cite{Sorkin} 
among them.

However, none of  modified space-time geometries 
has become generally accepted and able to replace the Minkowski geometry. 
Indeed, especial significance and reliability 
of the latter is stipulated by its origination from trustworthy 
physical principles of STR and, particularly, from the structure of 
experimentally verificated  Maxwell 
equations. None of its subsequent modifications can boast of such a firm 
and uniquely interpreted experimental ground.  

From the epoch of Minkowski we did not 
get better comprehension of the true geometry of our World, its hidden  
structure and origin. In fact, we are not even aware whether physical 
geometry is Riemannian or flat, has four dimensions or more, etc. Essentially, 
we can say nothing definite about the {\it topology\/} of space (both global 
and at a microscale). And, of course, we still have no satisfactory answer to 
the sacramental question: ``Why is the space three dimensional (at least, at 
a macrolevel)?'' Finally, an ``eternal'' question about the sense and origin 
of {\it physical time} stands as before on the agenda.

Meanwhile, the Minkowski geometry suffers itself from grave shortcomings 
both from phenomenological and generic viewpoints. To be concrete, 
complex structure of field equations accepted in quantum theory 
results, generally, in a string-like structure of field singularities 
(perhaps, this was first noticed by Dirac~\cite{Dirac}) and, 
moreover, these strings are unstable and, as a rule, radiate themselves 
to infinity (see, e.g.,~\cite{gensol} and the example in Section 2).

Another drawback (exactly, insufficiency) of the Minkowski geometry 
is the absence of fundamental distinction of temporal and 
spatial coordinates within its framework. Time enters the Minkowski metrical form on an equal  
footing with the ordinary coordinates though with opposite sign. In other words,  
in the framework of the STR geometry, time does not reveal itself as an  
{\it evolution parameter\/} which it had been, in fact, even in the antecedent 
Newton's picture of the World. At a pragmatic level this results, in particular,  
in the difficulty to coordinate ``times'' of various interacting 
(entangled) particles in an ensemble, 
in impossibility to introduce universal {\it global} time and to adjust the 
latter to {\it proper} times of different observers, or in the absence of clear 
comprehension of the passage of {\it local} time and dependence of its rate on  
matter. All these problems are widely discussed in physical literature 
(see, e.g.,~\cite{Horwitz}) but are still far from resolution.

However, the main discontent with generally accepted Minkowski geometry is  
related to the  fact that this geometry does not follow from some deep 
{\it logical} premises or exceptional {\it numerical} structures. This is 
still more valid with respect to generalizations of space-time structure  
arising, in particular, in the superstring theories (11D-spaces) and in 
other approaches for purely phenomenological, ``technical'' reasons which  
in no way can replace the transparent and general physical principles of STR: 
the principle of relativity and of the universal velocity of interaction 
propagation.   

At present, physics and mathematics are mature enough for construction of 
multidimensional geometries with different number of spatial and temporal 
dimensions. Moreover, they aim to create a general unified conception  
from which it would follow  definite conclusions on the {\it true} 
geometry of physical space and on the properties and meaning of physical  
time, on the {\it dynamics of Time\/} itself! 

In most of approaches of such kind the Minkowski space does not reproduce 
itself in its canonical form but is either deformed through some parameter 
(say, fundamental length and mass in the paradigm of Kadyshevsky  
~\cite{Kadysh}), under keeping correspondence with canonical scheme,  
or changes its structure in a more radical way. The latter takes place, in 
particular, in the theory of Euclidean time developed by  Pestov
~\cite{Pestov} (in this connection, see also~\cite{Montanus}), in the concept 
of Clifford space-time of Hestenes-Pavsic (see, e.g.,~\cite{Hestenes,Pavshich}), 
in  the framework of a special 6D geometry proposed by Urusovskii
~\cite{Urus}, etc.

At a still more fundamental level of consideration, one assumes to 
derive the geometry of physical space-time from some primordial 
principle encoding it (perhaps, together with physical dynamics).  
One can try to relate such an elementary {\it Code of Nature\/} with some 
exceptional symmetry 
(theory of physical structures of Kulakov~\cite{Kulakov} and binary 
geometrophysics of Vladimirov~\cite{Vlad2}), special group or algebra 
(quaternionic theory of relativity of Yefremov~\cite{Efrem} 
and algebrodynamics of Kassandrov~\cite{AD,Rev}), with an algebraically 
distinguished geometry (Finslerian anisotropic geometry of Bogoslovsky 
~\cite{Bogoslov} and geometry of polynumbers of Pavlov~\cite{Pavlov}) 
as well as with some special ``World function'' (metrical geometry of 
Rylov~\cite{Rylov}).  

Generally, all the above-mentioned and similar approaches affecting  
the very foundations of physics differ essentially one from another in the 
character of the first principle (being either purely physical or abstract in 
nature), in the degree of confidence of their authors to recently  
predominant paradigms (Lorentz invariance, Standard model, etc.) and in their  	
attitude towards the necessity to reproduce, in the framework of the original 
approach, the principal notions and mathematical insrumentation of the    
canonical schemes (Lagrangian formalism, quantization procedure, Minkowski 
space itself, etc.).In this respect the {\it neo-Pythagorean\/} philosophical paradigm professing 
by the author~\cite{number,code,Levich} seems most consistent and promising  
though difficult in realization.

Accordingly, under construction of an algebraic (logical, numerical) ``Theory  
of Everything'' one should {\it forget all of the known physical theories  
and even experimental facts\/} and to unprejudicely {\it read out} the 
laws of physical World in the internal properties of some exceptional 
abstract primordial structure, adding and changing nothing in the course of 
this for ``better correspondence with experiment''.  In this connection,  
one should be ready that physical picture of the World arising at the output 
could have little in common with recently accepted one and that 
{\it the real language of Nature} might be quite different from that we have 
thought out for a more effective description of the observable phenomena. 
In this situation
none {\it principle of correspondence} with former theories could be applied.

We have no opportunity to go into details of the neo-Pythagorean 
philosophy, quite novel and radical for modern science, sending the reader 
to Refs.~\cite{number,code,Levich}. Instead, in Section 2 we 
briefly present its realization in the framework of the ``old'' version of  
algebrodynamics developed during the period 1980-2005~\cite{AD,Rev}. Therein  
an attempt had been undertaken  to obtain the principal equations of physical 
fields and the properties of particle-like formations as the only consequence 
of the properties of the exceptional quaternion-like algebras, exactly, of 
the agebra of {\it biquaternions\/} $\mathbb{B}$.

We have forcible arguments to regard this attempt successful. From the 
sole  conditions of $\mathbb{B}$ analitivity (generalization of the 
Cauchy-Riemann equations, see Section 2) we were able to obtain 
an unexpectedly rich and rather realistic field theory.    
In particular, as a principal element of the arising picture of the World  
there turned to be a {\it flow of light-like rays} densely filling the space  
and  giving rise to a sort of particle-like formations at {\it caustics\/} 
and {\it focal points\/}. Such a primordial, matter generating structure 
has been called the {\it Flow of Prelight\/}. From mathematical point of 
view, this flow is defined by the {\it twistor structure\/} of the 
first equations for the biquaternionic field, whereas geometrically it 
represents itself a {\it congruence of null rays} of a special type 
(namely, {\it shear-free}), below -- the {\it generating congruence}. 

Meanwhile, the ``geometrical scene'' on which the algebraic dynamics displays 
itself has been, in fact ``by hands'', restricted to a subspace with 
canonical Minkowski metric for ensurence of the Lorentz invariance of the 
scheme. 
Such a procedure was in an evident contradiction with the whole philosophy 
of algebrodynamics, since corresponding subspace does not even form 
a subalgebra of $\mathbb{B}$ and is thus in no way  distinguished 
in the structure of the $\mathbb{B}$ algebra. From a more general viewpoint, 
neither in our old works nor in those of other authors there has  been 
found any {\it space-time algebra\/}, that is,  ascertained an algebraic  
(``numerical'') structure which could naturally induce the Minkowski 
geometry (or include the latter as its part)~\footnote{Hestenes was one of 
the first to consider the concept of space-time algebra~\cite{Hestenes}.  
We think, however, that his favourite 16D Dirac algebra cannot in fact be 
considered in this role since the additional dimensions have no natural 
physical interpretation.}. 

However, in 2005 in~\cite{Mink}  we have demonstrated that,  
under a thorough consideration, the primordial {\it complex\/} geometry of 
the $\B$ algebra unavoidably induces a {\it real\/} geometry just of the 
Minkowski type. In this scheme, the additional coordinates of (8D in reals) 
vector space of $\B$ are naturally compactified and behave like a 
{\it geometrical phase\/} suggesting thus a geometrical explanation of the  
{\it wave properties of matter\/} in general. In the following, this geometry 
has been called the {\it phase extension of the Minkowski space}. 
Its derivation and simplest properties  are presented in Section 3.   

Discovery of a novel geometry induced by the primordial algebraic struture of 
complex quaternions~\footnote{The algebra $\mathbb{B}$ is distinguished    
as a unification of the two exceptional (associative with norm and division) 
algebras, namely of the complex numbers and of Hamilton's quaternioins.},   
opens a wide perspective for the construction of a completed version of the  
algebrodynamics~\cite{PIRT05,Vestnik07}. In particular, it turns out that 
just (and only!) in the primordial complex space there can be realized one 
of the most interesting and original ideas of Wheeler-Feynman about ``all  
identical electrons as one and the same electron'', in its distinct positions  
on a unique world line. In~\cite{PIRT05}  a set of (dynamically correlated) 
copies of a sole  ``generating charge'' has been called {\it the 
ensemble of ``duplicons''}. We consider the geometrodynamical properties of 
duplicons and the related particle-like formations in Section 4.

In Section 5 a naturally arising concept of {\it complex time} 
is presented. Indeed, already in the previous version of the 
algebrodynamics (in the real Minkowski background) the temporal coordinate 
is distinguished in a natural way as an {\it evolution parameter} of the   
primordial biquaternionic (and of associated twistor) field: the generating 
Prelight Flow is then identified with the Flow of Time~\cite{number,Levich}. 
Now, in the complex 
pre-space such a parameter unavoidably turns to be two-dimensional, and the  
related order of the sequence of events -- indefinite. Thus in the framework 
of initially deterministic ``classical'' theory there arises inevitable 
{\it uncertainty of evolution of states\/} related to effectively stochastic 
alteration of the evolution parameter itself on the complex plane; we are 
led, therefore, to accept the concept of complex {\it random time}. On the 
other hand, existence of the geometrical phase makes it possible to suggest a 
novel treatment of the phenomena of {\it quantum interference}, alternative to  
the generally accepted concept of the {\it wave-particle dualism}. In particular, 
such a treatment relates the notion of the phase of wave function to  
the classical action of a particle, quite in the spirit of Feynman's 
version of quantum theory. Consideratons of these issues conclude the paper. 

\newpage

\section{Algebrodynamics over the Minkowski space-time}

Biquaternionic ($\B$) algebrodynamics is completely based on the  
(proposed by the author in 1980) version of noncommutative (including 
biquaternionic) analysis, that is, on the generalization of the theory   
of functions of complex variable to the case of noncommutative 
algebras of quaternionic type. This version is exposed in detail in the 
monograph~\cite{AD} (where one can find references to the preceding works) 
and in the recent review~\cite{Rev}.

Essentially, the whole structure of the theory of functions of $\B$ variable 
$Z\in \B$ follows from the invariant definition of a {\it differential} $dF$ of  
such, a {\it differentiable in $\B$}, function $F: Z \mapsto Z$ (a direct   
analog of an analytical function in the complex analysis). Specifically, 
in account of the associativity yet noncommutativity of the algebra, one has
\be{CRE}
dF = \Phi * dZ * \Psi,  
\ee
where $\Phi(Z),\Psi(Z)$ are some two auxiliary   
functions formerly called (left and right) {\it semi-derivatives} of $F(Z)$. 

Relation (\ref{CRE}) explicitly generalizes the well-known {\it Cauchy-Riemann 
conditions}. Indeed, in the case of the commutative algebra of complex numbers  
it acquires a familiar form
\be{CR}
dF = F^\p * dZ, 
\ee
with $F^\p:= \Phi * \Psi$ being the ordinary derivative of an analytical 
function $F(Z)$ of complex variable $Z$. Writing ({\ref{CR}) down in components,  
one comes to the standard Cauchy-Riemann set of equations. Thus, 
requirement of invariance of the differential (\ref{CR}) is a basic    
relation for constructing complex analysis, one of a number of equivalent 
well-known ones but  
suitable, moreover, for its generalization to the noncommutative case in the 
form (\ref{CRE}). Note that such a version was, perhaps, first proposed by 
Sheffers~\cite{Sheff} for construction of the analysis over an arbitrary  
commutative-associative algebra, and nearily after a century, was used by 
Vladimirov and Volovich~\cite{Volovich} for generalization to 
superalgebras.

Remarkably, in the case of real Hamilton quaternions  $\Q$ the proposed 
conditions (\ref{CRE}) reproduce another exceptional property of complex 
analysis, namely, the {\it conformity} of the correspondent mapping implemented  
by any analytical function~\cite{GR05,Acta}.   
However, since the conformal group of the Euclidean space ${\bf E}^k$ under 
$k \ge 3$ is finite (exactly, 15-parametrical for $k=4$), quaternionic analysis 
built on the base of relation (\ref{CRE}), turns to be unattractive with     
respect to the physical applications.

When, however, one passes to the case of $\B$ algebra (i.e., 
through the complexification of $\Q$), {\it the class of differentiable 
(in the sense of (\ref{CRE})) functions essentially expands} due to the 
existence of special elements of $\B$  --  {\it null divizors} 
(see details in~\cite{AD,Rev}; corresponding 
mappings have been called {\it degenerate conformal} ones). In this way we 
naturally arrive at the formulation of the first ``interpretational'' 
principle of the algebrodynamics:

\vskip2mm

\noindent
{\it In the paradigm of $\B$ algebrodynamics, there exists a unique 
fundamental physical field. This is a (essentially complicated and even 
multivalued) function of the $\B$ variable obeying the conditions of 
$\B$ differentiability (\ref{CRE}) --  the only primordial ``field 
equations''. All of the other ``fields'' arising in the scheme are 
secondary and can be defined through (semi)derivatives, contractions, etc. 
of the input $\B$ field. Their equations also follow from the 
``master equations'' (\ref{CRE}).}

\vskip2mm

Realization of this programme requires, meanwhile, the resolution of the 
problem of relationship between the 4D complex coordinate space $Z$, vector space of 
$\B$-algebra, and the Minkowski physical space-time. As it was already mentioned, 
the correct correspondence between these spaces has been ascertained only not 
long ago in our works and leads to a principally novel view on the geometry of space-time (see 
below). As to this section, we shall expose here only the former version   
of the algebrodynamics in which the coordinate space $Z$ is forcibly restricted 
onto the subspace with Minkowski metric, in order to guarantee the  
Lorentz invariance of the scheme and to avoid the problem of prescription  
of a particular meaning to the additional ``imaginary'' coordinates of the 
vector space of $\B$. 

Specifically, it is well known that the biquaternion algebra $\B$ is 
isomorphic to the full $2\times 2$ matrix algebra over $\C$.   
Further on we shall use the following two equivalent matrix representations   
of an element $Z\in \B$: 
\be{matrep}
Z =\begin{pmatrix}
u & w \\
p & v 
\end{pmatrix}
= \begin{pmatrix}
z_0 +z_3 & z_1 - iz_2 \\
z_1+iz_2 & z_0 - z_3 
\end{pmatrix}
\ee
through the four complex ``null'' $\{u,w,p,v\}$ or the ``Cartesian'' 
$\{z_\mu\},~\mu=0,1,2,3$ coordinates of a biquaternion $Z$ respectively.
Therefore, the 4D complex vector space of $\B$ possesses a natural complex 
(quasi)metrical form correspondent to the determinant $D$ of the 
representative matrix 
(\ref{matrep}), 
\be{determ}
D=z_0^2 - z_1^2 - z_2^2 -z_3^2.
\ee
This form turns into the Minkowski pseudo-Euclidean metric if only one 
considers the coordinates  $z_\mu \mapsto x_\mu \in \R$ as reals. This 
corresponds to the restriction of a generic matrix (\ref{matrep}) to a {\it 
Hermitean} one $Z\mapsto X=X^+$. However, we do not intend to restrict, in a 
similar way, the ``field'' matrices associated with {\it functions}  $F(X)$ 
of the space-time coordinates $X=\{x_\mu\}$, since the principal physical 
fields (especially in the framework of quantum theory) are generally 
considered as complex-valued. 

Thus we arrive at the second interpretational principle of the considered 
version of the algebrodynamics:

\vskip2mm

\noindent
{\it In $\B$ algebrodynamics, the coordinate physical space-time is represented  
by a subspace of the $4\C$ vector space of $\B$ correspondent to  
Hermitean $2\times 2$ matrices $X=X^+$ with their determinant representing j
ust the Minkowski metric. After such a restriction, the whole algebrodynamical 
scheme becomes manifestly Lorentz invariant.}

\vskip2mm

On the above described  coordinate ``cut'' correspondent to the Minkowski space, 
the initial conditions of $\B$ differentiability (\ref{CRE}) take the form 
\be{CREM}
dF = \Phi * dX * \Psi
\ee
and represent themselves a sort of ``master equations'' for some 
{\it algebraical field theory} uniquely determining all derivable 
properties of the latter. It is 
also noteworthy that none Lagrangians, commutation relations or other   
additional structures are used in the theory under consideration. Moreover: 
system of equations 
(\ref{CREM}) turns to be {\it overdetermined} and does not allow for any 
generating Lagrangian structure~\footnote{Lagrangian structure can be defined 
only for dynamical equations of secondary gauge fields associated with the 
``master equations'' (\ref{CREM}).}.

Consequently, the overdetermined character of the primordial algebrodynamical 
relations (\ref{CREM}), together with the {\it non-linearity} of the arising  
field equations (see below), makes it possible to consider the  
$\B$ algebrodynamics as a {\it theory of interacting fields} (and ``particles'',  
with rigidly fixed, ``self-quantized'' characteristics, see below).

As for particles, in the classical theory (like the algebrodynamics 
in its original form) in the capacity of those one can obviously take either 
regular (soliton-like) or singular field formations localized in the 3-space.  
Now we are ready to formulate the last (third) interpretational principle 
completing the set of first statements of the algebrodynamical scheme:

\vskip2mm

\noindent
{\it In the framework of $\B$ algebrodynamics, ``particles'' (particle-like 
formations) correspond to the (point- or extended but bounded in 3-space)  
singularities of the biquaternionic field, or to its derivatives. The latter  
may be put into correspondence with singularities of the secondary  
(Maxwell, Yang-Mills and other) fields which can be associated with any 
distribution of the primary $\B$ field. the shape, spatial arrangement, the 
characteristics and  temporal dynamics of these particle-like formations are 
again completely  determined by the properties of the master algebrodynamical 
system of equations for the $\B$ field (\ref{CREM}).}

\vskip2mm

It should be noted that symmetries (relativistic and conformal among them) 
of the system (\ref{CREM}) are considered in detail in the review~\cite{Rev}. 
The gauge and twistor structures specific for the system~(\ref{CREM}) 
are  also described therein (below we shall return to consideration of these). 
Let us now briefly review the principal properties and consequences of the 
$\B$ algebrodynamics (based solely on the conditions of $\B$ differentiability 
(\ref{CREM})).

\vskip2mm

1. Each {\it matrix} component $S(x,y,z,t)$ of a $\B$-differentiable function  
$F(X)$ satisfies the {\it complex eikonal equation}  
\be{eik}
(\frac{\prt S}{c \prt t})^2 -(\frac{\prt S}{\prt x})^2 - (\frac{\prt S}{\prt y})^2 - 
(\frac{\prt S}{\prt z})^2 = 0 .
\ee
This nonlinear, Lorentz and conformal invariant equation substitutes the 
Laplace equation in the complex analysis and form the basis of the algebraic 
field theory. 

2. Primary conditions (\ref{CREM}) can be reduced to a simpler set of 
equations of the form 
\be{spin}
d\xi = \Phi dX \xi
\ee 
for an effectively ``interacting''  2-spinor field  
$\xi(X)=\{\xi_A\},~A=1,2$ and potentials $\Phi(X)=\{\Phi_{AA^\p}\}$ of a 
complex gauge-like field (see for details~\cite{Rev}).  

3. Integrability conditions for the reduced overdetermined system 
({\ref{spin}) are just the {\it self-duality conditions} 
\be{selfdual}
F = iF^*
\ee
for the field strengths correspondent to the gauge potentials $\Phi_{AA^\p}$. 
Consequently, the complexified Maxwell and $SL(2,\C)$ Yang-Mills free equations 
are both satisfied on the solutions of the ``master system'' (\ref{spin}).

4. A field of a {\it null 4-vector} $k_\mu:~k^\mu k_\mu =0$ can be  
constructed from the fundamental 2-spinor $\xi(X)$ as follows: 
\be{isotrop}
k_\mu = \xi^+ \sigma_\mu \xi,
\ee
where $\sigma_\mu = \{I, \sigma_a\},~a=1,2,3$ is the canonical basis of 
the $2\times 2$ matrix algebra. As a consequence of ``master system'' 
(\ref{spin}), the null {\it congruence}  of rays tangent to $k_\mu$ is 
rectilinear (geodetic) and {\it shear-free}. This congruence of rays plays an 
extremely important role in the algebrodynamics; below we shall call it 
the {\it generating congruence}.   
~\footnote{Congruences like these naturally arise in the framework of the GTR 
and were widely studied, in particular, by Newman~\cite{Newman}, 
Kerr~\cite{KerrBur} and Burinskii~\cite{Burin}}. In the context of 
algebrodynamics it is important that an effective Riemannian metric 
$g_{\mu\nu}$ of a special form 
\be{riman}
g_{\mu\nu} = \eta_{\mu\nu} + h(X)k_\mu k_\nu
\ee
(the so-called Kerr-Schild metric~\cite{Schild}) may be put in correspondence 
with any $\B$ field or with associated generating congruence. This is a 
deformation of the flat Minkowski metric $\eta_{\mu\nu}$ preserving all the 
defining properties of generating congruence. Note that a self-consistent 
algebrodynamical scheme over a {\it curved\/} (algebraically special) 
space-time background has been developed in~\cite{Trish}.

5. In contrast to the ordinary nonlinear field models, in the algebrodynamics  
it turns out to be possible to obtain the {\it general solution} of the ``master 
system'' of equations (\ref{spin}) or (\ref{CREM}) in  an implicit 
{\it algebraic} form. The procedure is based upon the (well-known in the 
framwork of GTR) {\it Kerr theorem}~\cite{Schild,Penrose} that gives a full 
discription of the null shear-free congruences in the Minkowski or Kerr-Schild 
spaces, and makes also use of a natural generalization of this theorem~\cite{Eik,Pavlov04}, 
namely, of the general solution to the complex eikonal equation obtained 
therein. Briefly, the procedure of searching the solutions of eikonal 
equation and the associated congruence can be desribed as follows 
(for details see~\cite{Rev,Pavlov04}).  

Using gauge (projective) symmetry, one reduces the fundamental spinor 
$\xi(X)$ to the ratio of its two components choosing, say, 
\be{gauging}
\xi^T = (1,~g(X));
\ee
then {\it any solution of the algebrodynamical field theory is defined   
via the only complex function $g(x,y,z,t)$} -- the component of the 
{\it projective} 2-spinor $\xi$.

In turn, any solution for $g(X)$ is obtained in the following way. 
Consider an arbitrary (almost everywhere smooth) surface in the 
{\it 3D complex projective space $\C P^3$}; it may be set by an 
algebraic constraint of the form 
\be{kerr}
\Pi(g,\tau^1,\tau^2) = 0,
\ee
where $\Pi(...)$ is an arbitrary (holomorphic) function of three complex 
arguments. Let now these latters be linearily linked with the points of 
the Minkowski space through the so-called {\it incidence relation\/}~\cite{Penrose}
\be{inc}
\tau = X\xi
\ee
or, in components, 
\be{inc2}
\tau_1 = wg + u, ~~\tau_2 = vg + p,
\ee
where in the considered case of real Minkowski space the coordinates  
(\ref{matrep}) $u,v = ct \pm z$ are real and $p,w = x \pm iy$ 
complex conjugated. It is known that two spinors $\xi,\tau$  related with 
the points $X$ via the incidence relation (\ref{inc}) or  
(\ref{inc2}) form the so called {\it projective twistor} of the Minkowski space
~\cite{Penrose}. 

After substitution of (\ref{inc2}) into equation of generating surface 
(\ref{kerr}) the latter acquires the form of an algebraic equation   
\be{kerr2}
\Pi(g,wg+u,vg+p) = 0
\ee
with respect to the only unknown $g$, whereas the coordinates $\{u,v,p,w\}$ 
play the role of parameters. Resolving the equation above at  
each point of the Minkowski space $X$, one obtains some (generally 
multivalued) field distribution $g(X)$. 

In a rather puzzling way (the proof may be found, say, in~\cite{Wilson,Asya}), 
{\it for any generating function $\Pi$ and any continious branch of the 
solution under consideration, the field $g(X)$ identically satisfies 
both fundamental relativistic equations, the linear wave equation 
$\Box g = 0$ and the nonlinear equation of complex eikonal (\ref{eik})}.   
The correspondent spinor $\xi$ (in the gauge (\ref{gauging})) satisfies 
meanwhile the equations of shear-free null congruences and,  
according to the above-mentioned Kerr theorem, all such congruences can be 
obtained with the help of the exposed algebraic procedure.

6. It has been demonstrated in~\cite{Joseph,Trish99} that the  
complexified {\it electromagnetic field} associated with fundamental 
spinor $\xi$ which identically satisfies the self-duality conditions 
(\ref{selfdual} and thus the homogeneous Maxwell equations   
{\it can be directly expressed through the function} $g$ (obtained as a 
solution of the algebraic constraint (\ref{kerr2})) {\it and its derivatives} 
$\{\Pi_A, \Pi_{AB}\}$ {\it with respect to the twistor arguments}  
$\{\tau_A\},~A=1,2$. Specifically, for the {\it spintensor\/} of 
electromagnetic field strength $\varphi_{AB}$ one gets
\be{emstreng}
\varphi_{AB} = \frac{1}{P}\{\Pi_{AB} -\frac{d}{dg}(\frac{\Pi_A\Pi_B}{P})\},  
\ee
with $P:=d\Pi/dg$. The strengths of the associated Yang-Mills field can also be 
represented algebraically via (\ref{emstreng}) and the spinor $g$ itself.

7. It can be seen from representation (\ref{emstreng}) that the  
electroagnetic field strength turns to infinity at the points 
defined by the condition
\be{sing}
P=\frac{d\Pi}{dg} \equiv \frac{\prt \Pi}{\prt g} + w \Pi_1 + v \Pi_2 = 0.
\ee
Similar situation takes place for singularities of the associated Yang-Mills field 
and the {\it curvature field\/}  of the effective Kerr-Schild metric (\ref{riman}) 
(see~\cite{Schild,Wilson,Burin}). Therefore, in the context of    
$\B$ algebrodynamics one is brought to identify {\it particles with    
locus of common singularities of the curvature and gauge fields}. It is      
also reasonable to assume under this identification that, instantaneously,  
particle-like singularities are bounded in the 3-space
~\footnote{For string-like singularities expanding to infinity and found, 
e.g., in~\cite{Wilson})  another interpretation is needed (cosmic strings,  
etc.).}.  

8. With respect to the primary $\C$-field $g(X)$ obtained from 
the constraint (\ref{kerr2}) condition (\ref{sing}) 
defines its {\it branching points}. Geometrically, this  corresponds to 
{\it caustics} of the light-like rays of the generating congruence.
Generally speaking, instead of the primary $\B$ field and correspondent 
multivalued field $g(X)$, one can equivalently consider the fundamental 
congruence consisting, generically, of a (great) number of individual 
branches (``subcongruences''~\cite{Pavlov04}) and forming  caustics-particles 
at the points of merging of rays from some two of them, i.e. at the 
{\it envelope}. This all-matter-generating primordial structure in 
~\cite{number,Pavlov04,Levich} has been called the {\it pre-light flow}, 
or the ``Prelight''.  
 
9. At the same time, existence of the Prelight flow immediately distinguishes
the temporal structure~\footnote{Actually, this is true for any twistor structure  
in general.}. Indeed, the incidence relation (\ref{inc}) preserves its form 
under a one-parametrical coordinates transformation of the form 
\be{autM}
x_a \mapsto x_a + n_a s, ~~t \mapsto t + s,~~ n_a n_a = 1, ~~s \in \R , 
\ee
corresponding to a translation in the 3-space along each of rectilinear rays 
of the congruence, i.e. along the spatial directions specified by the unit 
vector $\bf n=\{n_a\}$,  
\begin{eqnarray}\label{dir}
{\bf n} = \frac{\xi^+ \bf \sigma \xi}{\xi^+\xi} = 
\qquad \qquad \qquad \nonumber \\
= \frac{1}{1+gg^*}\{g+g^*,~i(g-g^*),~ 1-gg^*\}.
\end{eqnarray}
Under such transformations physically correspondent to the process of 
{\it propagation} of the principal field with the universal velocity $V=c=1$,  
all of the three components of the projective twistor are preserved in 
value, as well as the direction vector $\bf n$ itself. Then, in accord with 
representation (\ref{autM}), one can regard these transformation as a 
prototype of the {\it course of time} and the Prelight Flow itself as the 
{\it Time Flow\/}. In more details these issues were considered in~\cite
{Pavlov04,Levich}, and in Section 5 we shall see in what an 
interesting way they are refracted under the introduction of a 
complex pre-space.

10. Particles identified above with common singularities of the gauge and 
curvature fields exhibit a number of remarkable properties specific for 
the real matter constituents. The most interesting is, perhaps, that of 
the {\it self-quantization} of electric charge. This property follows from  
the over-determinance of ``master system'' (\ref{spin}) and the self-duality 
of associated field strength (\ref{selfdual}). It is, at least partially, of 
topological origin. According to the quantization theorem proved in ~\cite{sing}), 
for any isolated and bounded (i.e., particle-like) singularity of 
the electromagnetic field (\ref{emstreng})  
electric charge is either null or necessarily integer multiple 
to some minimal, {\it elementary} value, namely, to the charge of fundamental 
static solution to the $\B$ equations (\ref{spin}). The latter is a  
direct analog of the well-known Kerr-Newman solution in the GTR. The solution 
follows from the twistor constraint (\ref{kerr}) with generating function 
$\Pi$ of the form 
\begin{eqnarray}\label{kerrsol}
\Pi = g\tau_1 - \tau_2 +2iag = \qquad \qquad \nonumber \\
= w g^2 + 2(z+ia)g -p =0,~~z:= \frac{u-v}{2}, 
\end{eqnarray}
resolving which, one obtains the two-valued solution 
\begin{eqnarray}\label{stereo}
g = \frac{p}{\hat z\pm \hat r} \equiv \frac{x+iy}{z+ia \pm 
\sqrt{x^2+y^2+(z+ia)^2}}, 
\end{eqnarray}
where $a=const \in \R$. With the above solution one can associate the famous 
{\it Kerr congruence} with the caustic locus of the form of a singular ring of 
radius $a$  correspondent to the locus of branching points of function 
(\ref{stereo}).  Particularly, in the degenerate case $a=0$ of a point-like 
singularity the associated via (\ref{emstreng}) electric field is the Coulomb 
one but the electric charge $q$ of singularity is strictly fixed in absolute 
value (in the accepted normalization $q=\pm 1/4$)~\cite{AD,GR05}. 
Correspondent effective 
metric (\ref{riman}) is just the Reissner-N\"ordstrem solution to the 
Einstein-Maxwell equations.

In the general case $a\ne 0$ solution (\ref{stereo}) leads to the field and 
metric exactly correspondent (under additional requirement on the electric 
charge to be unit!) to the above-mentioned Kerr-Newman solution (in the regime  
of a naked singularity free of horizon). B. Carter~\cite{Carter}) was the  
first to notice that correspondent gyromagnetic ratio for this 
field distribution is exactly equal to its anomalous value for the Dirac 
fermion. This observation stimulated subsequent studies (of Lopez, Israel, 
Burinskii, Newman et al.) in which the Kerr singular ring, with associated set 
of fields, has been regarded as a {\it model of electron\/}. Note that in the 	
algebrodynamical scheme this consideration is still more justified  
since the electric charge therein is necessarily fixed in modulus and 
may be identified as the elementary one. Thus,

\vskip2mm

\noindent
{\it in the framework of $\B$ algebrodynamics over Minkowski space the  
electron can be represented by the Kerr singular ring (of Compton size) 
related to a unique static axisymmetrical solution (\ref{stereo}) of equations  
(\ref{spin}), or of the constraint (\ref{kerr2}).}

\vskip2mm

11. A number of other exact solutions to the initial algebrodynamical 
equations and the related eikonal, Maxwell and Yang-Mills equations have  
been obtained in~\cite{Pavlov04,sing}, among them a {\it bisingular\/}   
solution and its toroidal modification~\cite{Asya}. They correspond to 
the case when generating function $\Pi$ in (\ref{kerr2}) is quadratic in $g$. 
More complicated solutions demand the computer assistance for solving 
the algebraic relation (\ref{kerr2}). However, the (most interesting) 
structure of singular loci of these distributions can be determined  
through elimination of the unknown $g$ from the set of two 
algebraic equations (\ref{kerr2}) and (\ref{sing})~\footnote{In the 
case of a polinomial form of the generating function $\Pi$ the procedure  
reduces to determination of the {\it resultant\/} of two polinoms  
and can be easily algorithmized}. The complex equation arising under 
the procedure, 
\be{eqm}
\Pi(x,y,z,t)=0
\ee
represents itself the {\it equation of motion} of particles-singularities and,  
moreover, at a fixed instant determines their spatial distribution and shape. 
In this way, we have examined the structure of singularities for 
a number of complicated solutions to ``master equations'' and equations of 
associated biquaternionic and electromagnetic fields. As for the latter, there 
has been obtained a peculiar solution to free Maxwell equations (!) 
describing the {\it process of annihilation\/} of two unlike 
(and necessarily unit) charges, with accompaning radiation of a  
singular wave front ~\cite{Pavlov04}, a class of the wave-like singular 
solutions~\cite{Trish99} etc.

11. If one restricts itself by {\it generic\/} solutions to ``master equations'' 
(\ref{spin}) or to associated Maxwell equations~\footnote{That is, by 
solutions free of any symmetry, in particular, being nonstatic and 
nonaxisymmetric.}, then their singular locus will (instantaneously) 
represent itself as {\it a number of one-dimensional curves -- ``strings''}. 
Generally, these strings (though neutral or carrying unit charges) are   
{\it unstable} in shape and size with respect to a small variation of 
parameters of the generating function $\Pi$. As an example, consider a  
special {\it deformation of the Kerr solution and congruence}~\cite{gensol}
defined by the following modification of the Kerr generating function: 
(\ref{kerrsol}):
\be{deform}
\Pi= g\tau^1(1-ih) -\tau^2(1+ih)+2iag, 
\ee
in which an additional parameter $h\in \R$ enters in addition to the 
standard Kerr parameter $a \in \R$. As a result, from the constraint 
$\Pi=0$ one obtains a novel solution for the function $g$ that defines  
still axisymmetrical but now time-dependent generating congruence of rays. 
Its caustic defined by the branching points of $g$ is represented by a 
uniformly collapsing into  
a point and, afterwards, expanding to infinity {\it singular ring\/}:
\be{defring}
z=0, ~~~\rho := \sqrt{x^2+y^2} = v(t-t_0),
\ee
where $t_0 = a/\sqrt{1+h^2}$ and velocity of collapse/expansion 
$v=h/\sqrt{1+h^2}$ is always less than the light one $c=1$. Thus, 

\vskip2mm

\noindent
{\it the Kerr congruence is unstable with respect to a small 
perturbation of controlling parameters of the generating function.
This let one expect also the instability of the Kerr (Kerr-Newman) 
solution of the (electro)vacuum Einstein equations, since the latter is   
defined, to a considerable degree, by the structure of a null congruence 
of the above-presented type.}

\vskip2mm

Note in addition that the deformed ring still carries a fixed, 
elementary charge but, nonetheless, is finally radiated to infinity.

\vskip2mm

At this point we complete our brief review of the ``old'' algebrodynamics
on the Minkowski background by the following remarks. In fact, from a 
single initial condition of $\B$ differentiability we were able to   
develop a self-consistent theory of fields and particle-like formations 
possessing a whole set of unique and physically realistic properties.  
The only {\it ad hoc} assumption made during the construction of the 
algebrodynamical theory, in order to ensure its Lorentz invariance, was a 
rather artificial restriction of the  
coordinate $4\C$ vector space of the $\B$ algebra onto the subspace with 
the Minkowski metrical form. On the other hand, the structure of 
{\it string-like\/} singularities-particles arising on $\bf M$ under 
this procedure turns out to be unstable and, perhaps, diffuses with time.  
Together, these considerations  suggest the necessity of a more successive 
analysis of the geometry ``hidden'' in the algebraic structure of  
biquaternion algebra, and of the probable links of its $4\C$ vector space
with the true physical geometry. On this way we immediately  
discover a completely novel geometry of the (extended) space-time 
presented in the next section.

\section{Biquaternion geometry and phase extension of the Minkowski space}

Let us return to matrix representation (\ref{matrep}) of the elements 
$Z\in \B$ of the biquaternion algebra. 
Restriction to {\it unitary} matrices $Z\mapsto U: U^+=U^{-1}*\det U$ reduces  
the algebra $\B$ to that of real Hamilton  quaternions $\Q$. Recall that $\Q$ 
is one of the two {\it exceptional associative division algebras}, 
together with the complex numbers algebra. The transformations preserving, 
together with the unitarity property, the structure of multiplication in $\Q$ 
(the inner automorphisms) are of the form     
\be{similar}
U \mapsto S*U*S^{-1},~~~S^{-1}=S^+ ,~S \in SU(2).
\ee
Under these, the diagonal (real) component of a matrix $U$ is invariant  
whereas the other three $\{x_1,x_2,x_3\}$ behave as the components of a 
rotating 3-vector (note that both $\pm S$ correspond to the same rotation: 
spinor structure). 
So the automorphism group of quaternion algebra 
$Aut(\Q)=SU(2)\cong SO(3)$ is $2:1$ isomorphic to the group of 3D rotations,  
with the main invariant 
\be{euclid}
l=x_1^2 + x_2^2 + x_3^2 ,
\ee
defining the {\it Euclidean\/} structure of geometry induced by the algebra 
$\Q$. In this sense, from the times of Hamilton, {\it exceptional algebra of  
real quaternions is considered as the algebra of physical background space}  
and, in the algebrodynamical paradigm, {\it predetermines its dimensionality 
and observable Euclidean structure}.

We can then apply the same ``Hamilton's logic'' to the algebra of 
biquaternions $\B$.  Now the elements $Z\in \B$ are represented by 
complex matrices of {\it generic\/} type (\ref{matrep}), and multiplication in 
$\B$ is preserved under the transformations 
\be{similarC}
Z \mapsto M*Z*M^{-1},~~~\det M = 1 ,~M \in SL(2,\C).
\ee
In full analogy with real case, the diagonal component $z_0$ in (\ref{matrep}) 
remains invariant, and the three others ${\bf z} = \{z_1,z_2,z_3\}$ 
manifest themselves as a 3D {\it complex\/} vector under {\it complex\/} 
rotations. Thus one has: $Aut(\B)=SL(2,\C)\cong SO(3,\C)$.  

Some explanations must be presented at this point. It is well known that the 6D (in reals) 
group $SL(2,\C)$ is a covering of the Lorentz group realizing its spinorial 
representation; the same is true for the $2:1$ isomorphic group of  
3D complex rotations $SO(3,\C)$. Specifically, Lorentz transformations  
can be represented in the form analogous to (\ref{similarC}),
\be{similarM}
X \mapsto M*X*M^{+},  
\ee
but act on the subspace $Z \mapsto X$ of {\it Hermitean\/} matrices 
$X=X^+$ with determinant representing the Minkowski metric. 
It is just this restriction that we considered in the previous 
section. Now, however, we are interested in a natural geometry induced by the 
{\it full} structure of the 8D (in reals) vector space $Z$ of the $\B$-algebra,  
in its hypothetical relationship to the Minkowski space and in the possible 
physical meaning of the four additional coordinates. It should be noted that, 
surprisingly, this geometry has not been discovered until now. As we shall see, 
corresponding construction is rather transparent and successive. 

We have seen that the structure of $\B$ multiplication is preserved under 
the 3D complex rotations forming the $SO(3,\C)$ group. The main {\it complex\/} 
invariant of these transformations, the analog of Euclidean invariant (\ref{euclid})  
of the real algebra $\Q$, is represented by  a {\it holomorphic\/}
(quasi)metrical bilinear form 
\be{quazimetr}
\sigma = z_1^2 + z_2^2 + z_3^2 \equiv  \vert {\bf z} \vert^2, 
\ee
the (squared) ``complex length'' of a vector $\bf z$.~It should be emphasized 
that all other metrical forms, the Hermitean metric among them, that could be 
canonically defined on the vector space $\C^4$ itself (or on its 
subspace $\C^3$) are, in fact, meaningless in the framework of the 
algebrodynamics since {\it they do not preserve their structure 
under the $\B$ automorphisms}.

On the other hand, from the complex invariant (\ref{quazimetr}) 
one can naturally extract a positive definite (exactly, non-negative) 
{\it Finslerian\/} metrical form of the 4-th degree by taking the square of 
complex modulus of the considered invariant 
\be{finsler}
S^2:=\sigma \sigma^* = \vert {\bf z} \vert^2   \vert {\bf z}^* \vert^2
\ee
As the next step, one can make use of the following remarkable identity
(see, e.g.,~\cite{Feudorov}):
\be{finsler2}
\vert {\bf z} \vert^2   \vert {\bf z}^* \vert^2 \equiv  ({\bf z} \cdot 
{\bf z}^*)^2  - \vert i~{\bf z} \times {\bf z}^* \vert^2
\ee                                   
that can be explicitly verified. 
Taking it into account, one can represent the positive-definite invariant      
(\ref{finsler}) in the form of a Minkowski-like interval~\cite{Mink}:
\be{Mink}
S^2 = T^2 - \vert {\bf R} \vert^2 \ge 0 , 
\ee
in which the quantities $T$ an $\bf R$ defined through the scalar $(\cdot)$ 
and vector $(\times)$ multiplications of complex 3-vectors as 
\be{effcoord}
T:= {\bf z} \cdot {\bf z}^*, ~~~{\bf R} : = i~{\bf z} \times {\bf z}^*,  
\ee 
aquire respectively the meaning of temporal and spatial coordinates of some 
effective 4D space with a Minkowski-type metric. 
Note also that such an identification is quite informal since under 
the $\B$ automorphisms acting as the 3D complex rotations the quantities 
$T,\bf R$ {\it transform one through the others just as the temporal and  
spatial coordinates do under the Lorentz transformations}~\footnote{
In fact, these transformations have some peculiarities in comparison  
with canonical Lorentz transformations, see~\cite{Mink} for details.}.

Thus, {\it the main real invariant of the biquaternion algebra, being 
positive definite, induces nonetheless the structure of causal    
domain of the Minkowski space} correspondent to the interior of 
the light cone (together with its light-like boundary). In this scheme, the 
events that are not causally connected as if do not exist at all (just as 
this should be from a successive viewpoint of the STR). We arrive, therefore, 
at a paradoxical but much interesting, both from physical and mathematical viewpoints, 
concept of the {\it physical space-time with a positive definite metric}.

Consider now the {\it phase} part of complex invariant 
(\ref{quazimetr}). The latter can be represented in the form 
\be{phase}
\sigma = S \exp^{i\alpha}, 
\ee
with absolute value $S$ correspondent to the Minkowski interval and the {\it 
phase} $\alpha$ also invariant under the 3D complex rotations (that is, in fact, 
under Lorentz transformations). In this connection, the  
{\it non-compact} (corresponding to modulus) part of the initial invariant is 
responsible for the {\it macro-geometry} explicitly fixed by an observer:  
remarkably, it turns to be exactly of the Minkowski type. At the same time, its 
phase, {\it compact} part determines geometry of the ``fiber'' and, perhaps, 
reveals itself at a {\it micro-level} being, in particular, related to  
the universal {\it wave} properties of matter (see Section 5). In the other 
respect, invariant $\alpha$ has the meaning of the {\it phase of the proper 
complex time\/} as this can be seen from (\ref{phase})  and will be 
discussed below.

We accept thus a novel concept of the background space-time geometry as of 
{\it the phase extension of a (causal part) of the Minkowski space\/} 
predetemined by the initial complex-quaternionic structure, with coordinates 
{\it bilinear\/} in those of the primordial and ``actually existing'' 
$\C^3$ space. 

Note that in literature dealing with various versions of complex extensions 
of the space-time geometry (see, e.g.,~\cite{Annals,Kechkin}) one usually 
encounters the procedure of separation of complex coordinates into real ``physical'' 
and imaginary ``unphysical'' parts (alternative to their separation into 
``modulus'' and ``phase'' parts in our approach). This procedure is, actually, 
inconsistent, since both parts are completely equivalent in their inner 
properties and should thus equally contribute to the induced real geometry one 
is going to construct.

Nonetheless, the above-mentioned {\it linear\/} separation of the complex 
coordinates is rather demonstrative. Specifically,     
consider {\it a couple of the 3D real vectors\/} $\{\bf p, \bf q\}$ associated 
with a complex vector $\bf z$ :
\be{pair}                                                          
{\bf z} = {\bf p} + i {\bf q}.
\ee
In this representation the principal invariant (\ref{quazimetr}) takes the form 
\be{sigma_pq}
\sigma = (\vert {\bf p} \vert^2 -\vert {\bf q} \vert^2) + i (2{\bf p} \cdot 
{\bf q}),
\ee
and corresponds to a pair of invariants in which one easily recognizes  
the two well-known {\it invariants of electromagnetic field} 
(with vectors $\bf p,\bf q$ identified as the field strengths  
of electric and magnetic field, respectively). The noticed analogy of 
complex coordinates and electromagnetic field strengths seems much suggesting  
and requires thorough analysis.

Express now, through the vectors $\bf p,\bf q$, the effective temporal and 
spatial coordinates (\ref{effcoord}):
\be{TX}
T = \vert {\bf p} \vert^2 + \vert {\bf q} \vert^2,~~~{\bf R} = 2{\bf p} \times 
{\bf q}  
\ee
and note that the temporal coordinate is positive definite 
(in Section 5 we shall relate this property with that of the 
{\it time irreversibility\/)}.  As to the three spatial coordinates, they form 
an {\it axial\/} vector so that the choice of sign corresponds to a reference 
frame of definite chirality. 

Finally, let us write down a remarkable relation~\cite{Mink} that links  
the module $V$ of the velocity ${\bf V} = \delta {\bf R} / \delta T$ of 
motion of a material point in the induced Minkowski space  with  
characteristics of the initial complex space $\C^3$, namely, with the invariant 
phase $\alpha$ and the {\it angle\/} $\theta$ between the vectors $\bf p$ and 
$\bf q$:
\be{connect}
\cos^2 \theta = \frac{1-V^2}{1+V^2 \coth^2 \alpha}.
\ee
In a limited case of motion with fundamental velocity $V=c=1$ one gets 
$\theta=\pi/2$, so that vectors $\bf p, \bf q$ are orthogonal to each other 
and to the direction of motion $\bf V$ (in analogy with an electromagnetic 
wave). From (\ref{TX}) one obtains also that in this ``light-like'' case  
the two vectors are equal in modulus, invariant $\sigma$ turns to zero, and the 
phase $\alpha$ becomes indefinite.

In the opposite case of a ``particle'' at rest $V=0$ one gets  
$\theta=0, \pi$, so that two different (``para'' and ``ortho'') relative 
orientations of vectors $\bf p, \bf q$ are possible. This remarkable property  
might be related to two admissible projections of the {\it spin vector\/} onto 
an abitrary direction in the 3-space.

\section{Complex algebrodynamics and the ensemble of ``duplicons''}

According to the first principles of the algebrodynamical approach, the true 
dynamics takes place just in the biquaternionic ``pre-space'' $\C^4$.  
In fact, we are able to explicitly observe only a ``shadow'' of this 
primordial dynamics on the induced (via mapping (\ref{effcoord})) 
real Minkowski-like space with the additional invariant phase and the causal 
structures.

As another ground for construction of {\it complex algebrodynamics\/} 
there serves a distinguished role of the ``complex null cone'', a direct 
analog of real Minkowski light cone. Specifically, consider two points 
$P,P^{(0)} \in \C^4$ with their coordinates $Z$ connected through the algebraic 
relation 
\begin{multline}\label{cone}
[z_1 - z_1^{(0)}]^2+[z_2 - z_2^{(0)}]^2+[z_3 - z_3^{(0)}]^2 = \\ 
\shoveright{= [z_0 - z_0^{(0)}]^2.}
\end{multline}
Then it is easy to demonstrate~\cite{PIRT05}, using the incidence relation  
$\tau=Z\xi$ (comp. with real case (\ref{inc})), that the twistor field  
(as well as the principal spinor field $g$ and the initial biquaternionic 
field) {\it takes equal values along the complex null line connecting the 
two points under consideration}, that is, along an {\it element\/} of the null 
cone. In this respect the positions and displacements of such points are 
correlated. It is noteworthy that in both sides of the {\it null cone equation\/} 
(\ref{cone}) there stands one of the two fundamental invariants of the $\B$ 
algebra so that {\it the primordial complex geometry dynamically reduces to the 
geometry of smaller space $\C^3$ with holomorphic (quasi)metrical form 
(\ref{quazimetr})}. As we are already aware of, this gives rise to a real 
effective geometry of the Minkowski type. Note also that for a fixed value of 
the (two equal) invariants equation (\ref{cone}) defines a complex 2-sphere. The latter manifold is 
$SO(3,\C)$-invariant and closely related to the unitary representations of 
the Lorentz group~\cite{Smorod}.

Finally, in full analogy with the ``old'' version of algebrodynamics on $\bf M$, 
let us {\it identify particles with singularities of the 
biquaternionic and the associated fields, geometrically -- with 
caustics of the generating congruence}. In this connection, recall that the 
generic type singularities on $\bf M$ have the structure of one-dimensional 
curves -- ``strings''. However, on the complexified $\C^3$ background 
singularities manifest a much richer structure.

Let us also emphasize from the beginning that the dynamical principles 
of complex algebrodynamics are completely the same as of its ``old'' 
version on $\bf M$. Specifically, we only make use of the  
biquaternionic fields obeying the $\B$-analyticity conditions or, 
equivalently, of twistor fields defining a {\it generating congruence\/} of 
complex null rays with zero shear. In particular, all the rules  
of definition of the set of relativistic fields (Section 2) do not 
require any modification in the complex case.

Consider now generating congruences (and, correspondingly, 
biquaternionic, twistor, and associated gauge fields) of a mathematically 
special and physically interesting type. These are congruences with a  
{\it focal line\/} -- a world line of some virtual point 
charge ``moving'' in the complex extension $\C^4$ of the Minkowski space
~\footnote{Exact definition and specification of such congruences 
is presented, say, in~\cite{PIRT05}.}. 
Structures like this have been first considered in the framework 
of GTR by Newman~\cite{Newman,Lind}; further on,   
congruences with a focal line will be called {\it Newman's congruences}
~\footnote{At present, this approach is intensively elaborating by   
Newman himself with collaborators~\cite{NewmanN} as well as 
by Burinskii~\cite{BurinN}.}.

In this connection, consider a point-like singularity-particle ``moving''  
in the complex space $\C^4$ along a ``trajectory'' $z_\mu = z_\mu(\tau), 
~\tau\in \C,~~\mu=0,1,2,3$. Points at which the primordial twistor (spinor) 
field takes the same values as in the vicinity of the ``particle'' are defined 
by the null cone equation (\ref{cone}). However, let these points belong themselves 
to the considered world line and represent thus other ``particles''. 
Then the null cone equation (\ref{cone}) acquires the form 
\begin{multline}\label{dubl}
L:=[z_1(\lambda) - z_1(\tau)]^2+[z_2(\lambda) - z_2(\tau)]^2+ \\
\shoveright{+ [z_3(\lambda) - z_3(\tau)]^2 -[z_0(\lambda) - z_0(\tau)]^2 = 0}
\end{multline}
and for any $\tau$ has, in general, a great (or even infinite) number of roots
$\lambda = \lambda^{(n)} (\tau)$ defining an ensemble of correspondent ``particles'' 
$z_\mu^{(n)}$. These are arranged in various points of the same complex  
world line and dynamically correlated (``interact'') with the initial 
(generating) particle -- the so-called {\it elementary observer}, see below.  

Such a set of ``copies'' of a sole pointlike particle ``observing itself''  
(both in its past and future, see Section 5) was first considered in our 
works~\cite{PIRT05,Vestnik07} and was called therein the ensemble of {\it duplicons\/}.
It is noteworthy that on the real Minkowski background equation  
(\ref{dubl}) (which on $\bf M$ turns out to be an ordinary 
{\it retardation equation}), in the case when the point of observation belongs 
itself to the world line of a particle, has a unique solution  
independently on the form of the trajectory, namely, the trivial solution  
$\lambda=\tau$.  Thus the concept of duplicons cannot be realized on the 
background of ordinary Minkowski space $\bf M$.

In~\cite{PIRT05,Vestnik07} we were guided by the old idea of 
R. Feynman and J.A. Wheeler~\footnote{Their  construction, by 
virtue of the above-mentioned reason, cannot be realized on real $\bf M$.} 
and considered each duplicon in the capacity of an {\it electron model}. Indeed, 
in full correspondence with the Feynman-Wheeler conjecture, in the arising 
picture ``all of the electrons'' are, essentially, ``one and the same electron''  
in various locations on a unique world line. In fact, however, the arising 
structure of singularities suggests a more natural, though exotic, 
interpretation.

Indeed, let us consider a primordial ``generating'' duplicon in the capacity of  
an ``elementary observer'' $\bf O$~\footnote{To model a real {\it macroscopic} 
observer, instead of a trajectory of an individual duplicon $z_\mu(\tau)$,  
one should introduce some {\it averaged} trajectory simplest of which 
is represented by a null complex line and, under its mapping into the   
real Minkowski space, corresponds to a uniform rectilinear motion of an 
inertial observer.}. All other duplicons on the null cone of the latter 
(\ref{dubl}), though dynamically correlated with $\bf O$, are in fact 
``invisible'' and not perceived by the elementary observer: 
{\it any ``signal'' is absent}! It is thus natural to conjecture that the 
act of ``perception'' (actually -- of {\it interaction}) takes place when only    
a null complex line (an element of the complex null cone) connecting $\bf O$ 
with some  duplicon becomes ``material'', that is, a {\it caustic} of 
the generating congruence.     
 
It is easy to determine the caustic locus of the congruence from the null cone 
equation (\ref{cone}). Similar to the case of general solution (\ref{kerr2})   
of the $\B$ analiticity equations, in the case of a Newman's congruence 
caustics coincide again with the {\it branching points} of the principal spinor 
field $g$ or, equivalently, of the field of {\it local time} of a duplicon     
$\tau(Z)$~\footnote{This field satisfies the complex eikonal equation
~\cite{PIRT05}. On the real Minkowski background this is analogous 
to the field of the ``retarded'' time.}. At these points one observes an  
{\it amplification} of the principal twistor-biquaternionic field 
(preserved along the elements of the complex null cone), and this can be 
regarded as the process of propagation of a ``signal'' to (from) 
the observer $\bf O$.    
 
In turn, branching points correspond to {\it multiple roots} of the 
null cone equation defined by the condition
\begin{multline}\label{mroots}
L^\p:=-\frac{1}{2}\frac{dL}{d\lambda} = z_a^\p (z_a(\lambda) - z_a(\tau))- \\
z_0^\p (z_0(\lambda) - z_0(\tau)) = 0\qquad 
\end{multline}
(summation over $a=1,2,3$ is assumed and prime denotes differentiation by 
$\lambda$). Together with the initial defining relation  
for duplicons (\ref{dubl}), the above condition specifies a {\it disrete set of 
positions of the observer (via its local times $\tau=\tau^{(k)}$) and 
of a pair of duplicons joining at a correspondent instant 
(defined as one of the mupliple roots $\lambda=\lambda^{(k)}$)}. Thus, 
an elementary {\it interaction act} can be regarded as a {\it merging of 
some two duplicons} ($a,b$) (with $\lambda(a)=\lambda(b)$) considered with 
respect to an observer $\bf O$ at some of his positions (with $\tau=\tau^{(k)}$). 
At such instants a process of the field amplification occurs along a null 
complex line, a caustic, connecting the observer and the two instantaneously 
coinciding duplicons. As it was already mentioned, under its mapping into 
$\bf M$ this line corresponds to some rectilinear path of a field 
{\it perturbation} moving in uniform with a velocity $V\le c$.    

We are now in a position to naturally distinguish particles-singularities 
arising in the scheme under consideration as the {\it matter constituents} and 
the {\it interaction carriers}, in full analogy with  the generally accepted  
theoretical classification.  
First of them form the ensemble of identical duplicons and can move along a 
very complicated and mutually concordant trajectories, geometrically -- along 
the {\it focal curve of the generating congruence}. As to the second, they 
always move along  rectilinear line elements of the complex null cone connecting 
a pair of ``interacting'' duplicons. In this process, the two merging duplicons 
represent an entire particle (see below) and stand for 
an {\it emitter} whereas the observer -- for a {\it detector} of the propagating 
``signal''; the problem of temporal ordering arising in this connection 
will be discussed in Section 5.

Thus, we are led to the conclusion that any {\it elementary object (electron?)} 
may be fixed by an observer only at some particular instants and 
{\it represents itself a pair of pre-elements, duplicons,  
emmiting a signal towards the observer when and only when their 
positions coincide in (complex) space}. At all the rest time these pre-elements -- duplicons 
-- are separated in space, do not radiate and, consequently, can be 
detected by none observer~\footnote{In the complex algebrodynamics 
there exists also another class of singularities representing themselves 
as a sort of ``three-element'' formations. One can speculate about their  
probable relationship to the the quark content of the matter}.

Conjecture about {\it duplicons as halves of the electron} revealing 
themselves solely at the instants of pairwise fusion strongly correlates with 
the modern concept and observations of {\it fractional charges\/} (see, e.g., 
the review~\cite{Bykov}) and, on the other hand, makes it possible to offer an    
alternative explanation for the wave properties of microobjects, particularly, 
for the {\it quantum interference phenomena\/}.

Indeed, let a pair of duplicons be identified as an electron 
at an instant of the first fusion, via the caustic-signal emitted towards an 
observer.  In the following, these ``twins'' diverge in space and, 
in particular, can pass through different ``slots'' in an idealized 
interference experiment. As a result, they  can again reveal themselves 
only at an instant of the next fusion accompanied by a new act of emission 
of a signal-caustic in the direction of the observer. Between the two 
fusions, each of the ``twins'' acquires a particular {\it phase lag\/}, 
namely, of the geometrical phase $\alpha$ of the principal complex invariant 
(\ref{phase}). Since, however, the complex coordinates of both ``twins'' 
at the instant of fusion should be equal, for the acquired phase lag one has    
\be{merge}
\Delta \alpha=2\pi N,~~ N=0,\pm 1, \pm 2, ~...~. 
\ee

Thus {\it there exists only a certain set of points at which a 
microobject might be once more observed after some its primary ``registration''}.
This stronly resembles the well- known procedure of {\it preparation} of a
quantum-mechanical state and of the following {\it QM measurement}, respectively. 
However, 
in the above presented picture we do not encounter any sort of the 
{\it wave-particle dualism\/}, of the de Broglie wave concept, etc. Each matter 
pre-element, duplicon, manifests itself as a typical pointlike 
corpuscular, whereas phase relations are of a completely geometrical nature and 
relate to some internal space of a ``fiber'' over $\bf M$~\footnote{In our 
scheme, the fiber itself defines the structure of the effective Minkowski base. 
Such situation is indeed unique and, in particular, can find application in the 
theory of the Calabi-Yau manifolds with a $3\C$ fiber structure.}. 
Below we shall once more return to discuss the interference phenomenon.

\section{Random complex time and quantum uncertainty}

Essentially, it is meaningless to discuss the problem of dynamics in complex 
space before one specifies the notion of {\it complex time\/}. In fact, we 
have already seen that the {\it evolution parameter\/}  
$\tau \in \C$ of an ``elementary observer'' $\bf O$, that is, the parameter 
of the ``world line'' $z_\mu (\tau)$ of a generating point singularity is now 
complex-valued. This means that the subsequent position of an observer on its 
``trajectory'' (under $\tau \mapsto \tau + d\tau$) is in fact {\it indefinite} 
by virtue of arbitrariness of alteration of the {\it phase} of the 
parameter $\tau$.

On the one hand, any value of $\tau$ one-to-one corresponds to a 
certain position of the observer $\bf O$ (and of the associated set of 
duplicons correlated with $\bf O$ through the null cone constraint) and, 
therefore, to {\it a definite ``state of the Universe''} with respect to a 
given observer.

On the other hand, a particular {\it realization\/} of those or other 
continuations of the trajectory is ambiguous being ruled by an unknown 
law of the ``walk'' of the evolution parameter $\tau$ across the complex plane, 
that is, by the form of a curve $\tau = \tau(t)$ with monotonically 
increasing real-valued parameter $t \in \R$~\footnote{One can evidently 
represent this parameter by the length of the curve.}. In~\cite{PIRT05,
Vestnik07} this was called the {\it evolution curve}.

Note that only after specification of the form of the evolution curve  
one can ascertain the {\it order} of succession of events and 
even distinguish past from future. It is just this form that defines 
the {\it time arrow\/} and predetermines, in particular, which of 
the (completely identical in dynamics)  duplicons is ``younger'' and 
which ``older'' than a certain ``elementary observer'' $\bf O$. . 

In the framework of the neo-Pythagorean ideology of algebrodynamics, 
the form of the universal evolution curve should follow from some 
general mathematical considerations and be exceptional with respect to its 
internal properties; unfortunately, at present the form is unknown. Up to now 
it only seems natural to expect that this ``Time Curve'' is extremely 
complicated and entangled (being, probably, of a fractal-like nature). 
Whether this is the case, {\it for us\/} the character of alteration of the 
evolution parameter on its complex plane would effectively represent 
itself a {\it random walk\/}. Moreover, one may conjecture that this 
walk is {\it discrete\/}, whereas the generating worldline $z_\mu (\tau)$
itself remains complex analytical: these two are completely independent. 
In the latter case, in the scheme there arises the {\it time quanta\/}, 
the ``chronon''. We shall see below that it has to be of the order of the 
Compton size, not of the Plank one. From different 
viewpoints the latter concept has been advocated in a number of works  
(see, e.g.,~\cite{Sidharth}  and references therein).  

It is noteworthy that, despite its probable random character, the Time Curve  
gives rise to mutually {\it correlated} alterations of the locations of 
different particles or, more generally, to {\it global synchronization of 
random processes of various nature}. At a microlevel, this may be 
related to the quantum nonlocality and entanglement, at a macrolevel -- 
to universal correlations already observed in the experiments of  
Shnoll (see, e.g., the review~\cite{Shnol}).  
                                        
Conjecture about random nature of the {\it Time dynamics\/} and resulting  
randomness of the motion of microobjects makes it possible to solve also 
the problem of concordance between the {\it increments\/} $\delta T,
\delta {\bf R}$ of the effective space-time coordinates (\ref{effcoord}) 
and the {\it differences\/} of their values for final and initial states   
$\Delta T,~\Delta {\bf R}$~\footnote{Generally, 
these are not necessarily equal due to the bilinearity of the induced space-time 
coordinates with respect to the primary complex ``holonomic'' coordinates 
$z_\mu$}. Indeed, say, for the time coordinate one gets  
\begin{multline}\label{deltat}
\Delta T: =  T^\p - T  =\\ 
= (\vert {\bf p} + d {\bf p} \vert^2 - \vert {\bf p} 
\vert^2) +  
+ (\vert {\bf q} + d {\bf q} \vert^2 - \vert {\bf q} \vert^2)  \\
= 2({\bf p} \cdot d {\bf p} + {\bf q} \cdot d {\bf q}) + 
(d {\bf p} \cdot d {\bf p} + d {\bf q} \cdot d {\bf q}). 
\end{multline}
Now under the {\it averaging\/} procedure the mixed term  
$dT = 2({\bf p} \cdot d {\bf p} + {\bf q} \cdot d {\bf q})$ vanishes, and the 
time interval 
\be{transtime}
\delta T = d {\bf p} \cdot d {\bf p} + d {\bf q} \cdot d {\bf q} 
\equiv \Delta T
\ee
at a ``physically infinitesimal'' scale behaves as a full differential, 
an actually holonomic entity. The same is true for the increments 
of {\it averaged\/} spatial coordinates $\delta {\bf R} \equiv 
\Delta {\bf R}$.

Moreover,  property of the increment of the time coordinate $\delta T \ge 0$ 
be positive definite ``in average'' leads immediately to a natural 
{\it kinematical\/} explanation of the {\it irreversibility\/} of physical time. 
Actually,  {\it any ``macroscopic'' alteration of the particles' positions 
(of the state of a system of particles) in the primary complex space necessarily 
results in an {\it increase\/} of the value of time coordinate of the effective 
Minkowski space}. Thus, in the algebrodynamical approach {\it irreversibility 
of time seems to be of kinematical and statistical nature}, and in the 
latter respect time resembles the {\it entropy\/}-like quantity in the 
orthodox scheme (if the latter is understood as a {\it probability measure\/}).

To conclude, in the context of initially {\it deterministic\/} ``classical'' 
dynamics there arises an unremovable {\it uncertainty\/}   
and, effectively, {\it randomness\/} of evolution of an observable ensemble of 
micro-objects. This uncertainty is of a global and universal character and is 
related to conjectural {\it stochastic\/} type of alteration of the 
complex time parameter, to {\it complex and effectively random nature of 
the physical time itself\/}. It is noteworthy that numerous 
problems and perspectives arising under introduction of the notion of 
{\it two-dimensional time\/} have been considered by Sakharov~\cite
{Sakharov};~Kechkin and Asadov~\cite{Kechkin} studied the 
quantum mechanics with complex time parameter and introduced, in this 
connection, the notion of different {\it alteration regimes} of this  
parameter similar to the above-introduced notion of the ``evolution curve''.

Remarkably, in the capacity of a rather unspecified parameter $\tau$ 
of the generating world line $z_\mu (\tau)$  one can (should!) use 
the principal invariant $\sigma$ of {\it complex proper time\/} (\ref{phase}), 
with its modulus $S$ corresponding to ordinary Minkowski proper time, and 
the phase $\alpha$ responsible for the uncertainty of evolution. This is the only  
parameter ensuring preservation of both the primary twistor 
field and the caustic structure (along straight null rays of 
the generating congruence) (see the proof in~\cite{PIRT05}). In this sense (despite its accepted 
name) complex ``proper'' time acquires the meaning of {\it universal 
global time\/} governing the concordant dynamics of the Universe.

We are now ready to return back to the analysis of quantum interference 
experiment started in the previous section. Recall that we have undertaken   
an attempt to relate the wave properties of matter to the conjecture of 
{\it dimerous electron} (formed by two pre-elements, duplicons, at the 
instants of their fusion) and to the geometrical phase (phase of the complex 
time $\alpha$) ``attached'' at each point of the generating world line 
and alterating along the latter. In the simplest case, assuming the linear 
proportionality of the (physically) infinitesimal increments of the modulus   
$dS$ and phase $d\alpha$ of complex time, 
\be{prop}
d\alpha = Const \cdot dS,
\ee
and choosing as the scale factor the inverse of a one-half of the Compton 
length $\lambda_0$ of the electron, $Const = (\lambda_0/2)^{-1} = 2Mc/\hbar$,  
(this corresponds to the above-mentioned assumption about the quanta of 
complex time, the ``chronon''), one obtains from the merging condition (\ref{merge})
\be{intrfr}
\Delta \alpha = \frac{2Mc}{\hbar} \Delta \int{dS}=\frac{\Delta A}{ \hbar} = 
2\pi N. 
\ee
Essentially, the above formula represents the condition for maxima of  
{\it classical interference\/} in the relativistic case. According to it, the 
phase lag for two ``halves''- duplicons under interference are proportional to 
their path difference, with the Minkowski interval as the invariant measure. On 
the one hand, this is in a remarkable correspondence with famous 
Feynmann representation of the {\it wave function\/} $\Psi=R\exp(iA/\hbar)$ 
whose phase is proportional to the classical action $A$ (for free 
particle -- to the proper time interval). On the other hand, 
in the nonrelativistic approximation, decomposing the interval $dS$ over the 
powers of velocity $V/c$ and taking into account the integrability of 
zero power term, one obtains as the {\it condition of quantum interference\/} 
the de Broglie relation 
\be{Broglie}
\Delta \int{\frac{dL}{\lambda}} = N,~~~\lambda:= \frac{h}{Mv},
\ee
with {\it integer\/} path difference of the two duplicons in fractions of the 
de Broglie wave length $\lambda$. 

Thus, the phase invariant $\alpha$ seems to be of fundamental physical 
importance being at the same time a measure of uncertainty of the 
evolution of micro-objects and the measure of their wave properties. The 
latters have their origin in the peculiarities of the primordial complex 
geometry and do not appeal to the paradigm of wave-particle dualism.

To conclude, we have endeavored to demonstrate the following.  
Exceptional complex geometry based on the properties of a remarkable algebraic 
structures (biquaternions), when introduced into foundations of physics 
as the primordial ``hidden'' geometry of the space-time (instead of the habitual 
Minkowski geometry), results in a quite novel and unexpected picture of 
the World. As its principal elements one can distinguish the identical 	 
pre-elements of matter -- duplicons -- constituents of observable particles 
(electrons?), as well as the uniformly propagating interaction carriers  
(caustics) and the effectively random complex time that predetermines the 
kinematical irreversibility of physical time at a macrolevel. However, after 25 years of 
development of the algebrodynamical field theory on ordinary Minkowski 
background, the ``new'' complex algebrodynamics is just at the very beginning 
of its march. We expect that the properties of biquaternion algebra and of 
the associated mathematical structures are rich enough to encode in themselves 
the  most fundamental laws of dynamics and geometry of the physical World.

\Acknow
{The author acknowledges S.I. Vinitsky and I.B. Pestov for fruitful 
discussions, A.V. Goosev for assistance in computer visualization of the 
duplicons' dynamics.}

\end{document}